\documentclass[preprintnumbers,amsmath,amssymb,floatfix,prd,nofootinbib,showkeys]{revtex4}

\usepackage{bm}
\usepackage{dcolumn}
\usepackage{graphicx}

\newcommand{\meV}{{\rm meV}}
\newcommand{\eV}{{\rm eV}}
\newcommand{\keV}{{\rm keV}}
\newcommand{\MeV}{{\rm MeV}}
\newcommand{\GeV}{{\rm GeV}}
\newcommand{\TeV}{{\rm TeV}}
\newcommand{\PeV}{{\rm PeV}}

\newcommand{\kpc}{{\rm kpc}}

\newcommand{\mG}{{\rm mG}}
\newcommand{\s}{{\rm s}}

\newcommand{\Mpl}{M_{\rm Pl}}

\begin{document}
\preprint{DESY 09-149}




\title{Lorentz Violation: Motivation and new constraints}


\author{Stefano Liberati}\email{liberati@sissa.it} \affiliation{SISSA, Via Beirut 2-4, 34014 Trieste (Italy) \\INFN, Sezione di Trieste, Via Valerio 2, I-34127 Trieste (Italy)} 
\author{ Luca Maccione} \email{luca.maccione@desy.de} \affiliation{DESY, Theory Group, Notkestrasse 85, D-22607 Hamburg (Germany)}
\date{\today} 
\keywords{
Quantum Gravity phenomenology, special relativity, high-energy astrophysics, Cosmic-Rays, Planck scale, Crab Nebula}

\begin{abstract}
We review the main theoretical motivations and observational constraints on Planck scale suppressed violations of Lorentz invariance. After introducing the problems related to the phenomenological study of quantum gravitational effects, we discuss the main theoretical frameworks within which possible departures from Lorentz invariance can be described. In particular, we focus on the framework of Effective Field Theory, describing several possible ways of including Lorentz violation therein and discussing their theoretical viability. We review the main low energy effects that are expected in this framework. We discuss the current observational constraints on such a framework, focusing on those achievable through high-energy astrophysics observations. In this context we present a summary of the most recent and strongest constraints on QED with Lorentz violating non-renormalizable operators. Finally, we discuss the present status of the field and its future perspectives.
\end{abstract}

\maketitle

\section{Introduction}

Physics is an observational and experimental science; it is about the ingenious interrogation of Nature and the interpretation of its answers via a mathematical language. Although this may seem an obvious statement, it is nonetheless an increasingly harder point to maintain as theoretical physics proceeds toward more abstract and contrived issues. However, without the guidance of experiments and observations, all our theories are doomed to remain mathematical constructs without connection to reality.
The most striking such case is the decades old quest for a quantum theory of gravitation.  

Quantum Gravity (QG) has posed a challenge to many theoretical physicists of the last generation and it remains far from understood. Although we do not yet have a single experiment or observation forcing us to introduce such a theory\footnote{However, part of the gravitation theory community would say that current cosmological observations (dark energy and dark matter issues) are definitely taking up this role.}, we definitely need it, not only on philosophical grounds (reductionism as a driving force in physics), but also because we know that in physically relevant regimes (e.g.~singularities in cosmology and in black holes...) our classical theory of gravitation, i.e.~General Relativity (GR), fails to be predictive. However, when searching for QG, we have to confront not only deep theoretical problems (e.g.~the renormalizability of gravitational theories, the possible loss of unitarity in gravitational phenomena \cite{Hawking:1976ra}, the meaning of time in QG \cite{Isham:1995wr,Butterfield:1998dd})  but also the lack of observational and experimental guidance. The typical scale at which QG effects should become relevant is expected to be the one at which the gravitational action (the Einstein-Hilbert action for GR) becomes of the order of the quantum of action $\hbar$. This happens at the so called Planck scale $\Mpl \equiv \sqrt{\hbar c/G_{N}}\simeq 1.22\times 10^{19}~\GeV/c^{2}$ which corresponds to energies well above the capabilities of any Earth based experiment as well as any observationally accessible regime.

However, it was realized (mainly over the course of the past decade) that the situation may not be as bleak as it appears. In fact, models of gravitation beyond GR and models of QG have shown that there can be several of what we term low energy ``relic signatures'' of these models, which would lead to deviation from the standard theory predictions (standard model of particle interactions (SM) plus GR) in specific regimes. Some of these new phenomena, which comprise what is often termed ``QG phenomenology'', include: 
\begin{itemize}
\item Quantum decoherence and state collapse \cite{Mavromatos:2004sz}
\item QG imprint on initial cosmological perturbations \cite{Weinberg:2005vy}
\item Cosmological variation of couplings \cite{Damour:1994zq,Barrow:1997qh}
\item TeV Black Holes, related to extra-dimensions \cite{Bleicher:2001kh}
\item Violation of discrete symmetries \cite{Kostelecky:2003fs}
\item Violation of space-time symmetries \cite{Mattingly:2005re}
\end{itemize}
In this review we focus upon the phenomenology of violations of fundamental symmetries, given that a convenient way to perform high-precision tests is to look for experimental deviations from symmetries that are believed to hold {\em exactly} in nature and that could be broken by QG. 

An example of such a fundamental symmetry is CPT invariance, which requires that physics be unchanged under the combination of charge conjugation (C), parity inversion (P) and time reversal (T). C connects particles and antiparticles, P represents a spatial reflection of physical quantities with respect to the coordinate origin and T reverses a physics reaction in time.

In Quantum Field Theory, Lorentz symmetry is intimately related to CPT symmetry. Indeed, one of the hypotheses of the well known ``CPT theorem'' is Lorentz invariance. If CPT is broken, then at least one of the hypotheses of the CPT theorem should also break down. It has been proven \cite{Greenberg:2002uu} that Lorentz symmetry is the failing assumption in the so called ``anti-CPT theorem'', which states that in any unitary, local, relativistic point-particle field theory CPT breaking implies Lorentz violation. Note however that the converse of this statement is not true: it is possible to violate Lorentz invariance while keeping CPT exact\footnote{However, this theorem does not hold for theories that do not admit a field-theoretic description and that can therefore have unexpected properties.}.

Thus, it is interesting to study both the theory and the phenomenology of Lorentz invariance violation (LV), which may yield a glimpse of QG. Although from the theoretical point of view the exploration of this possibility has been  active for many years \cite{DiracAet,Bj,Phillips66,Blokh66,Pavl67,Redei67}, a phenomenology of LV has developed only within the last ten years or so. Before the mid-1990s, few works investigated the experimental consequences of LV (see however \cite{1978NuPhB.141..153N,1980NuPhB.176...61E,Zee:1981sy,Nielsen:1982kx,1983NuPhB.217..125C,1983NuPhB.211..269N,1984NuPhB.242..542.}), because new effects were expected only in particle interactions at energies of order the Planck mass $\Mpl$. Later, it was realised that there are special situations in which new effects could manifest also at lower energy. These situations were termed ``Windows on Quantum Gravity''.

\section{Windows on Quantum Gravity}

In recent years, attempts to place constraints on high-energy deviations from LI have mainly focused on modified dispersion relations for elementary particles.  Indeed, specific hints of LV arose from various approaches to Quantum Gravity. Among the many examples are string theory tensor VEVs \cite{KS89}, space-time foam~\cite{AmelinoCamelia:1997gz}, semiclassical spin-network calculations in Loop QG~\cite{Gambini:1998it}, non-commutative geometry~\cite{Carroll:2001ws, Lukierski:1993wx, AmelinoCamelia:1999pm}, some brane-world backgrounds~\cite{Burgess:2002tb} and condensed matter analogues of ``emergent gravity''~\cite{Barcelo:2005fc}. 

In most of the above mentioned QG models, LV enters through modified dispersion relations. These relations can be cast in the general form\footnote{We assume that rotational invariance is preserved and that only boost invariance is affected by Planck-scale corrections (see \cite{Jacobson:2005bg} for a discussion about this assumption).}
\begin{equation}%
E^2=p^2+m^2+f(E,p;\mu;M)\;,%
\label{eq:disprel}%
\end{equation}%
where he low energy speed of light $c=1$; $E$ and $p$ are the particle energy and momentum, respectively; $\mu$ is a particle-physics mass-scale (possibly associated with a symmetry breaking/emergence scale) and $M$ denotes the relevant QG scale. Generally, it is assumed that $M$ is of order the Planck mass: $M \sim M_{\rm Pl} \approx 1.22\times 10^{19}\;$GeV, corresponding to a quantum (or emergent) gravity
effect. The function $f(E,p;\mu;M)$ can be expanded in powers of the momentum (energy and momentum are basically indistinguishable at high energies, although they are both taken to be smaller than the Planck scale), and the lowest order LV terms ($p$, $p^2$ and $p^3$) have primarily been considered \cite{Mattingly:2005re}\footnote{We disregard  the possible appearance of dissipative terms \cite{Parentani:2007uq} in the dispersion relation, as this would correspond to a theory with unitarity loss and to a more radical departure from standard physics than that envisaged in the framework discussed herein.}.

At first glance, it appears hopeless to search for effects suppressed by the Planck scale. Even the most energetic particles ever detected (Ultra High Energy Cosmic Rays, see, e.g.,~\cite{Roth:2007in,Abbasi:2007sv}) have $E \lesssim 10^{11}~\GeV \sim 10^{-8} \Mpl$. 
However, even tiny corrections can be magnified into a significant effect when dealing with high energies (but still well below the Planck scale), long distances of signal propagation, or peculiar reactions (see, e.g.,~\cite{Mattingly:2005re}).
A partial list of these {\em windows on QG} includes:
\begin{itemize}
\item sidereal variation of LV couplings as the lab moves
  with respect to a preferred frame or direction
\item cumulative effects: long baseline dispersion and vacuum birefringence (e.g.~of signals from gamma ray bursts, active galactic nuclei, pulsars)
\item anomalous  (normally forbidden) threshold reactions allowed by LV terms (e.g.~photon decay, vacuum
\v{C}erenkov effect) 
\item shifting of existing threshold reactions (e.g.~photon annihilation from
  blazars, GZK reaction)
\item LV induced decays not characterised by a
threshold (e.g.~decay of a particle from one helicity to the other
or photon splitting)
\item maximum velocity (e.g.~synchrotron peak from supernova
remnants)
\item dynamical effects of LV background fields (e.g.
  gravitational coupling and additional wave modes)
\end{itemize}

However not all of these tests are similarly robust against the underlying physical framework that one chooses to justify the use of the modified dispersion relations of the form (\ref{eq:disprel}). Although the above cited cumulative effects exclusively use the form of the modified dispersion relations, all the others depend on the underlying dynamics of interacting particles and on whether or not the standard energy-momentum conservation holds. Thus, to cast most of the constraints on dispersion relations of the form (\ref{eq:disprel}), one needs to adopt a specific theoretical framework justifying  the use of such deformed dispersion relations.

\section{Theoretical models}

Although many kinematic frameworks have been proposed to account for LV, dynamically meaningful realisations of LV are more interesting from a phenomenological point of view, as they provide a more complete framework in which to compute reactions. An obvious requirement for any such model is that it agree with known experimental observations. Thus, a convenient way to study LV is to embed it into an effective framework that contains the SM by construction. 

Effective Field Theory (EFT) is a well established means of describing physics. It is particularly suitable for our purposes because it provides a sufficiently robust and general set of rules to describe LV physics without requiring that we know the details of the QG models leading to such effects. Indeed, many QG models can be reduced to EFT with LV (see e.g.~\cite{Gambini:1998it,Carroll:2001ws,Burgess:2002tb}). Moreover, it is widely believed that the SM itself, although it can describe particle interactions up to at least ${\rm few}\times 100~\GeV$ at unprecedented precision, could be such an effective model. 

In EFT, the presence of non-renormalizable operators in the Lagrangian is a signal of the presence of new higher energy interactions. One can infer the energy scale at which these operators are generated (and at which the theory breaks down) by simply looking at their dimensionfull coupling constants and their associated mass scale. (A typical example is $4-$fermion interaction, with the coupling constant $G_{F}\propto M_{W}^{-2}$.) 
On the contrary, for the SM this type of reasoning cannot be applied because its renormalizability does not allow one even to guess at which scale new physics should appear.

When constructing an EFT of Planck-scale LV, one can deal with this restriction in two ways. On the one hand, one can introduce LV terms in the SM Lagrangian by adding only operators that preserve renormalizability. On the other hand, one can  explicitly break Lorentz invariance by introducing non-renormalizable operators of mass dimension larger than four. The most crucial difference between the two approaches is that the renormalizable operators lead to contributions that are relevant even at low energy (and hence generically need very suppressed couplings to fit experimental data), but the non-renormalizable operators are relevant only at high-energy and, moreover, are naturally Planck-scale suppressed. Both possibilities have been explored in the so called Standard Model Extension (SME), which we discuss in the following section.

\subsection{The Standard Model Extension with renormalizable operators}

Most of the research in EFT performed with only renormalizable ({\em i.e.}~mass dimension 3 and 4) LV operators has been carried out within the so called (minimal) SME~\cite{KS89}. It consists of the SM plus all LV renormalizable operators ({\em i.e.}~of mass dimension $\leq4$) that can be written without changing the field content or violating gauge symmetry. The operators appearing in the SME can be conveniently classified according to their behaviour under CPT \cite{Mattingly:2005re}. 

Because the most common particles used to cast constraints on LV are photons and electrons, a prominent role is played by LV QED. The high energy ($\Mpl\gg E \gg m$) dispersion relations for QED can be expressed as (see \cite{Mattingly:2005re} and references therein for more details)
\begin{eqnarray} \label{eq:SMErotinvdisp}
E_{e}^2=m_{e}^2+p^2+f^{(1)}_e p+f^{(2)}_ep^2\\
E_{\gamma}^2=(1+ f^{(2)}_\gamma ){p^2}
\end{eqnarray}
where the first equation is for electrons and the second one is for photons. The coefficients $f_{e}^{(1)}$, $f^{(2)}_{e}$ and $f^{(2)}_{\gamma}$ depend in general on the helicity state of the particles and are related to the coupling parameters of the LV operators in the Lagrangian \cite{Mattingly:2005re}. 
The positron dispersion relation is the same as (\ref{eq:SMErotinvdisp}) with the replacement $p\rightarrow -p$, which only changes the $f^{(1)}_e$ term.

Note that the typical energy, $p_{\rm crit}$, at which a new phenomenology should begin to appear is quite low. If, for example $f_{e}^{(2)} \sim O(1)$, the corresponding extra term is comparable to the electron mass $m$ precisely at $p_{\rm crit} \simeq m_{e} \simeq 511~\keV$. Even worse, for the linear modification to the dispersion relation, in the case where $f^{(1)}_{e} \simeq O(1)$, we find that $p_{\rm crit} \sim m^{2}/\Mpl \sim 10^{-17}~\eV$. (Note that, by chance, this energy is close to the present upper limit on the photon mass, $m_{\gamma}\lesssim 10^{-18}~\eV$ \cite{pdg}.) 

However, the natural values for the parameters $f_{e}^{(n)}$ may be much lower than $O(1)$. For example, they can be suppressed by ratios of $(m_{e}/\Mpl)^{\delta}$, where $\delta>0$. If we take $\delta=1$, then the suppression factor is $m_{e}/\Mpl\simeq 4\times10^{-23}$, which is not too far from the limits that have been placed on dimension 4 LV parameters to date.

Because a rich literature is available on constraints on the minimal SME \cite{Kostelecky:2008zz}, we focus on non-renormalizable extensions in the following section.

\subsection{The Standard Model Extension with non-renormalizable operators}

The lowest order non-renormalizable LV operators for SME have mass dimension 5. Myers \& Pospelov \cite{Myers:2003fd} found that there are essentially only three operators of dimension 5, quadratic in the fields, that can be added to the QED Lagrangian and that preserve rotation and gauge invariance, but break local LI\footnote{Actually, these criteria allow the addition of other (CPT even) terms, but these would not lead to modified dispersion relations (they can be thought of as extra, Planck suppressed, interaction terms) \cite{Bolokhov:2007yc}.}.

These operators, which result in a contribution of $O(E/\Mpl)$ to the dispersion relation of the particles, are:
\begin{equation}
-\frac{\xi}{2\Mpl}u^mF_{ma}(u\cdot\partial)(u_n\tilde{F}^{na}) + \frac{1}{2\Mpl}u^m\bar{\psi}\gamma_m(\zeta_1+\zeta_2\gamma_5)(u\cdot\partial)^2\psi\:,
\label{eq:LVterms}
\end{equation}
where $\tilde{F}$ is the dual of $F$ and where $\xi$ and $\zeta_{1,2}$ are dimensionless parameters. All these terms also violate CPT symmetry.
Recently, this construction was extended to the whole SM \cite{Bolokhov:2007yc}. 

From (\ref{eq:LVterms}) the dispersion relations of the fields can be
modified as follows. For the photon, we find
\begin{equation}
\omega_{\pm}^2 = k^2 \pm \frac{\xi}{\Mpl}k^3\:,
\label{eq:disp_rel_phot}
\end{equation}
where the $+$ and $-$ signs denote right and left circular polarisation, respectively. For the fermion, we find 
\begin{equation}
E_\pm^2 = p^2 + m^2 + \eta_\pm \frac{p^3}{\Mpl}\;,
\label{eq:disp_rel_ferm}
\end{equation}
where $\eta_\pm=2(\zeta_1\pm \zeta_2)$and where the $+$ and $-$ signs denote positive and
negative helicity states, respectively.  For the antifermion, simple ``hole interpretation" arguments show that the same dispersion relation holds, yielding $\eta^{af}_\pm = -\eta^f_\mp$ where
$af$ and $f$ denote anti-fermion and
fermion coefficients, respectively~\cite{Jacobson:2005bg,Jacobson:2003bn}.

Observations involving very high energies can thus potentially cast an $O(1)$ constraint on the coefficients defined above.  A natural question arises: what is the theoretically expected value of the LV coefficients in the modified dispersion relations shown above?  

This question is clearly intimately related to the meaning of any constraint procedure. Indeed, let us suppose that, for an unknown reason, the dimensionless coefficients $\eta$, which according to the well-known Dirac criterion should be of order $O(1)$, are defined up to a dimensionless factor of $m_{e}/\Mpl \sim 10^{-22}$.  Then, any constraint of order $\gtrsim 10^{-22}$ would be ineffective, assuming that our aim is to learn something about the underlying QG theory. 

This problem could be further exacerbated by renormalization group effects, which in principle could strongly suppress the low-energy values of the LV coefficients even if they are $O(1)$ at high energies. However, the renormalization group equations for the LV parameters for dimension 5 LV QED have been derived in \cite{Bolokhov:2007yc} and they show that the running is logarithmic. Therefore, low energy constraints are robust against renormalization group effects.

In conclusion, because we lack a definite QG model to rely on, we assume our LV parameters to be $O(1)$ at the Planck scale and we judge the strength of our constraints against this reference value.

\subsection{The naturalness problem and higher dimension LV operators} 
\label{sec:naturalness}

There is however a problem with the theory described above. It is indeed generic that even starting with an EFT with only LV of mass dimension 5 and 6 for free particles, radiative corrections due to particle interactions will generate lower-dimension LV terms that will then become dominant~\cite{Collins:2004bp}, as their dimensionless coefficients are of the same order as the higher dimension ones ($O(1)$, given our previous assumption, see \cite{Mattingly:2008pw}). Thus, radiative corrections do not preserve a dispersion relation of the form shown in  (\ref{eq:disp_rel_phot},\ref{eq:disp_rel_ferm}), but they automatically induce extra unsuppressed LV terms in $p$ and $p^2$, which dominate the $p^3$. 
Either a symmetry (or some other mechanism) protects the lower dimension operators from large LV, or the suppression of the non-renormalizable operators indeed will always be greater than that of the renormalizable ones. 

A possible solution to this problem is provided by SuperSymmetry (SUSY) \cite{GrootNibbelink:2004za,Bolokhov:2005cj}, a symmetry relating fermions to bosons, i.e.~matter with interaction carriers. SUSY is intimately related to Lorentz invariance. Indeed, it can be shown that the composition of at least two SUSY transformations induces space-time translations. However, SUSY can still be an exact symmetry even in the presence of LV and it can serve as a custodial symmetry, preventing certain operators from appearing in LV field theories. 

The effect of SUSY on LV is to prevent dimension $\leq 4$, renormalizable LV operators to be present in the Lagrangian.
Moreover, it has been demonstrated \cite{GrootNibbelink:2004za,Bolokhov:2005cj} that the renormalization group equations for Supersymmetric QED plus the addition of dimension 5 LV operators \`a la Myers \& Pospelov do not generate lower dimensional operators, if SUSY is unbroken. However, this is not the case for our low energy world, of which SUSY is definitely not a symmetry. 
The effect of soft SUSY breaking was again investigated in \cite{GrootNibbelink:2004za,Bolokhov:2005cj}. As expected, it was found that, when SUSY is broken, the renormalizable operators appear in the Lagrangian. In particular, dimension $\kappa$ operators arise from the percolation of dimension $\kappa+2$ LV operators\footnote{We consider  only $\kappa = 3,4$, for which these relationships have been demonstrated.}. The effect of SUSY soft-breaking is, however, to introduce a suppression of order $m_{s}^{2}/\Mpl$ ($\kappa=3$) or $(m_{s}/\Mpl)^{2}$ ($\kappa=4$), where $m_{s}$ is the scale of soft SUSY breaking. Given the present constraints, the theory in which $\kappa=3$ must be fine-tuned to be viable, because the SUSY-breaking-induced suppression is not powerful enough to eliminate linear modifications in the dispersion relation of electrons. However, if $\kappa = 4$, the induced dimension 4 terms are sufficiently suppressed, provided that $m_{s} < 100~\TeV$.

In summary, mass dimension 5 LV operators seem to be unnatural, even considering the effects of SUSY, because the corresponding LV parameters have to be much less than their ``natural'' value $O(1)$ to fit current low energy data. However, dimension 6 LV does not suffer from this problem. There is no clear argument accounting for why the dimension 5 operators should not appear in the high energy theory \footnote{One could argue that the action for gravity in GR is proportional to $G_{N}\sim \Mpl^{-2}$; therefore the leading order contributions have to be suppressed by 2 powers of $\Mpl$ and hence must have dimension 6. However, the Liouville inspired string theory model \cite{Ellis:1999jf,Ellis:2003sd} is an example in which the leading order LV terms have mass dimension 5 (as the string action is proportional to $\sqrt{G_{N}}$) \cite{Sarkar:2002mg}.}. However, it can be shown that if we assume CPT invariance for the Planck scale theory, then dimension 5, CPT odd LV operators are forbidden, and only dimension 6 operators can appear.

Therefore, CPT and (soft broken) SUSY produce a viable LV theory.
This is encouraging enough to allow us to consider this theory a serious candidate for LV, but currently no conclusive statements can be made.

For the reasons given above, it is interesting to study theories with higher dimension contributions. The candidate theory should preserve CPT and should be supersymmetric. In the absence of a dynamical model, we can proceed effectively by adding to the SM (actually, for simplicity, to the QED) all possible dimension 6, CPT even operators \cite{Mattingly:2008pw}. 


The complete dimension 6 SME is not known. We still lack the LV induced interaction terms and the CPT odd kinetic ones. This is not a severe limitation, however. Indeed, LV induced interactions are expected to have a very suppressed rate; thus, we do not expect them to be observable in elementary particle experiments. On the other hand, we have already shown that the major attraction of dimension 6 SME is essentially related to the assumption that CPT is an exact symmetry; therefore, we neglect CPT odd terms.

The CPT even dimension 6 LV terms have only recently been computed \cite{Mattingly:2008pw} through the same procedure used by Myers \& Pospelov for dimension 5 LV.
The known nonrenormalizable CPT even fermion operators are
\begin{eqnarray}
&\nonumber -\frac{1}{\Mpl}\bar{\psi}(u\cdot D)^{2}(\alpha_{L}^{(5)}P_{L}+\alpha_{R}^{(5)}P_{R})\psi \\
& - \frac{i}{\Mpl^{2}}\bar{\psi}(u\cdot D)^{3}(u\cdot \gamma)(\alpha_{L}^{(6)}P_{L} + \alpha_{R}^{(6)}P_{R}) \psi \\
& \nonumber-\frac{i}{\Mpl^{2}}\bar{\psi} (u\cdot D) \square (u\cdot \gamma) (\tilde{\alpha}_{L}^{(6)}P_{L} + \tilde{\alpha}_{R}^{(6)}P_{R}) \psi\;,
\label{eq:op-dim6-ferm}
\end{eqnarray}
where $P_{R,L}$ are the usual left and right spin projectors $P_{R,L} = (1\pm\gamma^{5})/2$ and where $D$ is the usual QED covariant derivative. All coefficients $\alpha$ are dimensionless because we factorize suitable powers of the Planck mass.

The known photon operator is
\begin{equation}
-\frac{1}{2\Mpl^{2}}\beta_{\gamma}^{(6)}F^{\mu\nu}u_{\mu}u^{\sigma}(u\cdot\partial)F_{\sigma\nu}\;.
\label{eq:op-dim6-phot}
\end{equation}

From these operators, the dispersion relations of electrons and photons can be computed, yielding
\begin{eqnarray}
\nonumber 
E^{2} - p^{2} - m^{2} &=& \frac{m}{\Mpl}(\alpha_{R}^{(5)}+\alpha_{L}^{(5)})E^{2} + \alpha_{R}^{(5)}\alpha_{L}^{(5)}\frac{E^{4}}{\Mpl^{2}} \\ 
& & + \frac{\alpha_{R}^{(6)} E^{3}}{\Mpl^{2}}(E+sp) + \frac{\alpha_{L}^{(6)}E^{3}}{\Mpl^{2}}(E-sp) \\
\nonumber 
\omega^{2}-k^{2} &=& \beta^{(6)}\frac{k^{4}}{\Mpl^{2}}\;,
\label{eq:disp-rel-dim6}
\end{eqnarray}
where $m$ is the electron mass and where $s = \bm{\sigma}\cdot\mathbf{p}/|\mathbf{p}|$. Also, notice that a term proportional to $E^{2}$ is generated. However, this term is suppressed by the tiny ratio $m/\Mpl \sim 10^{-22}$ and can be safely neglected, provided that $E > \sqrt{m\Mpl}$.
 
Because the high-energy fermion states are almost exactly chiral, we can further simplify the fermion dispersion relation in eq.~(\ref{eq:disp-rel-dim6}) (we pose $R=+$, $L=-$)
\begin{equation}
E^{2} = p^{2} + m^{2} + f_{\pm}^{(4)} p^{2} + f_{\pm}^{(6)} \frac{p^{4}}{\Mpl^{2}}\;.
\label{eq:disp-rel-ferm-dim6-improved}
\end{equation}
Because it is suppressed by $m/\Mpl$, we will drop in the following the quadratic contribution $f_{\pm}^{(4)}p^{2}$ \cite{Mattingly:2008pw}\footnote{This is an example of a dimension 4 LV term, generated at high energy, with a natural suppression of $m_{e}/\Mpl \sim 10^{-22}$. Therefore, any limit larger than $10^{-22}$ placed on this term does not have to be considered as an effective constraint. To date, the best constraint for electron LV parameters of dimension 4 in SME is $O(10^{-17})$ \cite{Kostelecky:2008ts}.}.

It may seem puzzling that in a CPT invariant theory we distinguish between different fermion helicities. However, although they are CPT invariant, some of the LV terms displayed in eq.~(\ref{eq:op-dim6-phot}) are odd under P and T. However, CPT invariance allows us to determine a relationship between the LV coefficients of the electrons and those of the positrons. Indeed, to obtain it we must consider that, by CPT, the dispersion relation of the positron is given by (\ref{eq:disp-rel-dim6}), with the replacements $s \rightarrow -s$ and $p\rightarrow -p$. This implies that the relevant positron coefficients $f^{(6)}_{\rm positron}$ are such that $f^{(6)}_{e^{+}_{\pm}} = f^{(6)}_{e^{-}_{\mp}}$, where $e^{+}_{\pm}$ indicates a positron of positive/negative helicity (and similarly for the $e^{-}_{\pm}$).

\subsection{Other frameworks}
\label{sec:others}

Altough EFT is a natural choice of framework in which to study LV, there are other possibilities, arising in some models of string theory, that deserve attention. Indeed, if in the high energy theory a hidden sector exists that cannot be accessed because it lives, for example, on a different $D-$brane from us, there are LV effects that cannot be fit in an EFT description. Because the EFT approach is nothing more than a highly reasonable, but rather arbitrary ``assumption'', it is worth studying and constraining additional models, given that they may evade the majority of the constraints discussed in this review.

\subsubsection{D-brane models}
\label{sec:nonEFT}

We consider here the model presented in \cite{Ellis:1999jf,Ellis:2003sd}, in which modified dispersion relations are found based on the Liouville string approach to quantum space-time \cite{Ellis:1992eh}. Liouville-string models of space-time foam \cite{Ellis:1992eh} motivate corrections to the usual relativistic dispersion relations that are first order in the particle energies and that correspond to a vacuum refractive index $\eta = 1-(E/\Mpl)^{\alpha}$, where $\alpha = 1$. Models with quadratic dependences of the vacuum refractive index on energy: $\alpha = 2$ have also been considered \cite{Burgess:2002tb}.

In particular, the D-particle realization of the Liouville.string approach predicts that only gauge bosons such as photons, not charged matter particles such as electrons, might have QG-modified dispersion relations. This difference may be traced to the facts that \cite{Ellis:2003if} excitations which are charged under the gauge group are represented by open strings with their ends attached to the D-brane \cite{Polchinski:1996na}, and that only neutral excitations are allowed to propagate in the bulk space transverse to the brane. Thus, if we consider photons and electrons, in this model the parameter $\eta$ is forced to be null, whereas $\xi$ is free to vary. Even more importantly, the theory is CPT even, implying that vacuum is not birefringent for photons ($\xi_{+} = \xi_{-}$).

\subsubsection{Doubly Special Relativity}

Lorentz invariance of physical laws relies on only few assumptions: the principle of relativity, stating the equivalence of physical laws for non-accelerated observers, isotropy (no preferred direction) and homogeneity (no preferred location) of space-time, and a notion of precausality, requiring that the time ordering of co-local events in one reference frame be preserved \cite{ignatowsky,ignatowsky1,ignatowsky2,ignatowsky3,ignatowsky4,Liberati:2001sd,Sonego:2008iu}. 

All the realizations of LV we have discussed so far explicitly violate the principle of relativity by introducing a preferred reference frame. This may seem a high price to pay to include QG effects in low energy physics. For this reason, it is worth exploring an alternative possibility that keeps the relativity principle but that relaxes one or more of the above postulates. Such a possibility can lead to the so-called very special relativity framework \cite{Cohen:2006ky}, which was discovered to correspond to the break down of isotropy and to be described by a Finslerian-type geometry~\cite{Bogoslovsky:2005cs,Bogoslovsky:2005gs,gibbons:081701}. In this example, however, the generators of the new relativity group number fewer than the usual ten associated with Poincar\'e invariance. Specifically, there is an explicit breaking of the $O(3)$ group associated with rotational invariance. 

One may wonder whether there exist alternative relativity groups with the same number of generators as special relativity. Currently, we know of no such generalization in coordinate space. However, it has been suggested that, at least in momentum space, such a generalization is possible, and it was termed  ``doubly" or ``deformed" (to stress the fact that it still has 10 generators) special relativity, DSR.  Even though DSR aims at consistently including the dynamics, a complete formulation capable of doing so is still missing, and present attempts face major problems. Thus, at present DSR is only a kinematic theory. Nevertheless, it is attractive because it does not postulate the existence of a preferred frame, but rather deforms the usual concept of Lorentz invariance in the following sense.

Consider the Lorentz algebra of the generators of rotations, $L_i$,
and boosts, $B_i$:%
\begin{equation}%
[L_i,L_j] = \imath\,\epsilon_{ijk}\; L_k\;; \qquad%
[L_i,B_j] = \imath\,\epsilon_{ijk}\; B_k\;; \qquad%
[B_i,B_j] = -\imath\,\epsilon_{ijk}\; L_k%
\label{LB}
\end{equation}%
(Latin indices $i,j,\ldots$ run from 1 to 3) and supplement it with
the following commutators between the Lorentz generators and those of
translations in spacetime (the momentum operators $P_0$ and $P_i$):%
\begin{equation}%
[L_i,P_0]=0\;; \qquad%
[L_i, P_j] = \imath\,\epsilon_{ijk}\; P_k\;;%
\label{LP}%
\end{equation}%
\begin{equation}%
[B_i,P_0] = \imath\; f_1\left(\frac{P}{\kappa}\right) P_i\;;%
\label{BP0}%
\end{equation}%
\begin{equation}%
[B_i,P_j] = \imath\left[ \delta_{ij} \; f_2\left(\frac{P}{\kappa}\right)
P_0 + f_3\left(\frac{P}{\kappa}\right)  \frac{P_i \; P_j}{\kappa}
\right]\;.%
\label{BPj}%
\end{equation}%
where $\kappa$ is some unknown energy scale.
Finally, assume $[P_i,P_j] = 0$. %
%
%
The commutation relations (\ref{BP0})--(\ref{BPj}) are given in terms
of three unspecified, dimensionless structure functions $f_1$, $f_2$,
and $f_3$, and they are sufficiently general to include all known DSR
proposals --- the DSR1~\cite{AmelinoCamelia:2000mn}, DSR2~\cite{Magueijo:2001cr,Magueijo:2002am}, and DSR3~\cite{AmelinoCamelia:2002gv}.  Furthermore, in all the DSRs considered to date, the dimensionless arguments of these functions are specialized to%
\begin{equation}%
f_i\left(\frac{P}{\kappa}\right)  \to f_i\left( \frac{P_0}{\kappa},
\frac{\sum_{i=1}^3 P_i^2}{\kappa^2}\right)\;,%
\end{equation}%
so rotational symmetry is completely unaffected.  For
the $\kappa\to +\infty$ limit to reduce to ordinary special relativity, $f_1$ and $f_2$ must tend to $1$, and $f_3$ must tend to some finite value.%

DSR theory postulates that the Lorentz group still generates space-time symmetries but that it acts in a non-linear way on the fields, such that not only is the speed of light $c$ an invariant quantity, but also that there is a new invariant momentum scale $\kappa$ which is usually taken to be of the order of $\Mpl$. Note that DSR-like features are found in models of non-commutative geometry, in particular in the $\kappa$-Minkowski framework \cite{Lukierski:2004jw,AmelinoCamelia:2008qg}, as well as in non-canonically non commutative field theories \cite{Carmona:2009ra}.

Concerning phenomenology, an important point about DSR in momentum space is that in all three of its formulations (DSR1~\cite{AmelinoCamelia:2000mn}, DSR2~\cite{Magueijo:2001cr,Magueijo:2002am}, and
DSR3~\cite{AmelinoCamelia:2002gv}) the component of the four momentum having deformed commutation with the boost generator can always be rewritten as a non-linear combination of some energy-momentum vector that transforms linearly under the Lorentz group \cite{Judes:2002bw}. For example in the case of DSR2~\cite{Magueijo:2001cr,Magueijo:2002am} one can write
s%
\begin{eqnarray}%
&&E=\frac{-\pi_0}{1-\pi_0/\kappa}\;;%
\label{msE}\\%
&&p_i=\frac{\pi_i}{1-\pi_0/\kappa}\;.%
\label{msp}%
\end{eqnarray}%
It is easy to ensure that while $\pi$ satisfies the usual dispersion
relation $\pi_0^2-\mbox{\boldmath $\pi$}^2=m^2$ (for a particle with
mass $m$), $E$ and $p_i$ satisfy the modified relation%
\begin{equation}%
\left(1-m^2/\kappa^2\right)E^2+2\,\kappa^{-1}\,m^2\,E
-\mbox{\boldmath $p$}^2=m^2\;.%
\label{mod-disp}%
\end{equation}%
Furthermore, a different composition for energy-momentum now holds, given that the composition for the physical DSR momentum $p$ must be derived from the standard energy-momentum conservation of the pseudo-variable $\pi$ and in general implies non-linear terms.

Despite its appealing, DSR is riddled with many open problems.
First,  if DSR is formulated as described above --- that is, only in momentum space --- then it is an incomplete theory.  Moreover, because it is always possible to introduce the new variables $\pi_\mu$, on which the
Lorentz group acts in a linear manner, the only way that DSR can avoid
triviality is if there is some physical way of distinguishing the
pseudo-energy $\epsilon\equiv -\pi_0$ from the true-energy $E$, and
the pseudo-momentum $\mbox{\boldmath $\pi$}$ from the true-momentum
$\mbox{\boldmath $p$}$. If not, DSR is no more than a nonlinear choice of coordinates in momentum space.%

In view of the standard relations $E\leftrightarrow \imath\hbar
\partial_t$ and $\mbox{\boldmath $p$}\leftrightarrow -\imath\hbar
\mbox{\boldmath $\nabla$}$ (which are presumably modified in DSR), it is clear that to physically
distinguish the pseudo-energy $\epsilon$ from the true-energy $E$, and
the pseudo-momentum $\mbox{\boldmath $\pi$}$ from the true-momentum
$\mbox{\boldmath $p$}$, one must know how to
relate momenta to position. At a minimum, one needs to develop
a notion of DSR spacetime.%

In this endeavor, there have been two distinct lines of approach, one
presuming commutative spacetime coordinates, the other attempting to
relate the DSR feature in momentum space to a non commutative position
space. In both cases, several authors have pointed out major
problems. In the case of commutative spacetime coordinates, some
analyses have led authors to question the triviality~\cite{Ahluwalia:2002wf} or
internal consistency~\cite{Rembielinski:2002ic,Schutzhold:2003yp} of DSR.  On the other hand, non-commutative proposals \cite{Lukierski:1993wx} are not yet well
understood, although intense research in this direction is under way\cite{AmelinoCamelia:2007wk}. 

DSR is still a subject of active research and debate \cite{Smolin:2008hd,Rovelli:2008cj}; nonetheless, it has not yet attained the level of maturity needed to cast robust constraints\footnote{However, some knowledge of DSR phenomenology can be obtained by considering that, as in Special Relativity, any phenomenon that implies the existence of a preferred reference frame is forbidden. Thus, the detection of such a phenomenon would imply the falsification of both special and doubly-special relativity. An example of such a process is the decay of a massless particle.}. For these reasons, in the next section we focus upon EFT and discuss the constraints within this framework.%


\section{Constraints on Effective Field Theory with $O(E/\Mpl)$ Lorentz invariance violation}

There is now a well-established literature on the experimental and observational constraints on renormalizable operators. Given that such operators involve effects that are not {\it a priori} Planck suppressed, experimental tests indeed constrain the coefficients of dimension 3 and 4 LV operators to be very small \cite{Kostelecky:2008zz}. Caveat the above discussion on the natural values of such coefficients, we focus here on the lowest order non-renormalizable operators. Let us begin with a brief review of the most common types of constraints.

For definiteness, we refer to the following modified dispersion relations:
\begin{eqnarray}
\label{eq:mdr}
E^{2}_{\gamma} &=& k^{2} + \xi_{\pm}^{(n)}\frac{k^{n}}{\Mpl^{n-2}}\\
E^{2}_{el} &=& m_{e}^{2} + p^{2} + \eta_{\pm}^{(n)}\frac{p^{n}}{\Mpl^{n-2}}\;,
\end{eqnarray}
where, in the EFT case, we impose $\xi^{(n)} \equiv \xi_{+}^{(n)} = (-)^{n}\xi_{-}^{(n)}$ and $\eta^{(n)} \equiv \eta_{+}^{(n)} = (-)^{n}\eta_{-}^{(n)}$.

\subsection{Photon time of flight}
\label{subsec:tof}

Although photon time-of-flight constraints currently provide limits several orders of magnitude weaker than the best ones, they have been widely adopted in the astrophysical community. Furthermore they were the first to be proposed in the seminal paper \cite{AmelinoCamelia:1997gz}.  More importantly, given their purely kinematical nature, they may be applied to a broad class of frameworks beyond EFT with LV. For this reason, we provide a general description of time-of-flight effects, elaborating on their application to the case of EFT below.

In general, a photon dispersion relation in the form of (\ref{eq:mdr}) implies that photons of different colors (wave vectors $k_1$ and $k_2$) travel at slightly different speeds. 

Let us first assume that there are no birefringent effects, so that $\xi_{+}^{(n)} = \xi_{-}^{(n)}$.
Then, upon propagation on a cosmological distance $d$, the effect of energy dependence of the photon group velocity produces a time delay
\begin{equation}
 \Delta t^{(n)} = \frac{n-1}{2}\, \frac{k_2^{n-2}-k_1^{n-2}}{\Mpl^{n-2}}\,\xi^{(n)}\, d\;,
\label{eq:tof-naive}
\end{equation}
which clearly increases with $d$ and with the energy difference as long as $n>2$. The largest systematic error affecting this method is the uncertainty about whether photons of different energy are produced simultaneously in the source.

So far, the most robust constraints on $\xi^{(3)}$, derived from time of flight differences, have been obtained within the $D-$brane model (discussed in section \ref{sec:nonEFT}) from a statistical analysis applied to the arrival times of sharp features in the intensity at different energies from a large sample of GRBs with known redshifts~\cite{Ellis:2005wr}, leading to limits $\xi^{(3)}\leq O(10^3)$.
A recent example illustrating the importance of systematic uncertainties can be found in \cite{Albert:2007qk}, where the strongest limit $\xi^{(3)} < 47$ is found by looking at a very strong flare in the TeV band of the AGN Markarian 501. 

One way to alleviate systematic uncertainties (available only in the context of birefringent theories, such as the one with $n=3$ in EFT) would be to measure the velocity difference between the two polarisation states at a single energy, corresponding to
\begin{equation}
 \Delta t = 2|\xi^{(3)}|k\, d/\Mpl\;.
\end{equation}
This bound would require that both polarisations be observed and that no spurious helicity-dependent mechanism (such as, for example, propagation through a birefringent medium) affect the relative propagation of the two polarisation states.

However, eq.~(\ref{eq:tof-naive}) is not valid when applied to birefringent theories. Photon beams generally are not circularly polarized; thus, they are a superposition of fast and slow modes. Therefore, the net effect of this superposition may partially or completely erase the time-delay effect. To compute this effect on a generic photon beam in a birefringent theory, let us describe a beam of light by means of the associated electric field, and let us assume that this beam has been generated with a Gaussian width
\begin{equation}
\vec{E} = A\, \left(e^{i(\Omega_{0}t-k^{+}(\Omega_{0})z)}\,e^{-(z-v_{g}^{+}t)^{2}\delta\Omega_{0}^{2}}\hat{e}_{+} + e^{i(\Omega_{0}t-k^{-}(\Omega_{0})z)}\,e^{-(z-v_{g}^{-}t)^{2}\delta\Omega_{0}^{2}}\hat{e}_{-}       \right)\;,
\end{equation}
where $\Omega_{0}$ is the wave frequency, $\delta\Omega_{0}$ is the gaussian width of the wave, $k^{\pm}(\Omega_{0})$ is the ``momentum'' corresponding to the given frequency according to (\ref{eq:mdr}) and $\hat{e}_{\pm}\equiv (\hat{e}_{1}\pm i\hat{e}_{2})/\sqrt{2}$ are the helicity eigenstates. Note that by complex conjugation $\hat{e}_{+}^{*} = \hat{e}_{-}$. Also, note that $k^{\pm}(\omega) = \omega \mp \xi \omega^{2}/\Mpl$. Thus,
\begin{equation}
\vec{E} = A\, e^{i\Omega_{0}(t-z)}\left(e^{i\xi\Omega_{0}^{2}/\Mpl z}\,e^{-(z-v_{g}^{+}t)^{2}\delta\Omega_{0}^{2}}\hat{e}_{+} + e^{-i\xi\Omega_{0}^{2}/\Mpl z}\,e^{-(z-v_{g}^{-}t)^{2}\delta\Omega_{0}^{2}}\hat{e}_{-}       \right)\;.
\end{equation}
The intensity of the wave beam can be computed as
\begin{eqnarray}
\nonumber \vec{E}\cdot\vec{E}^{*} &=& |A|^{2}\left( e^{2i\xi\Omega_{0}^{2}/\Mpl z} + e^{-2i\xi\Omega_{0}^{2}/\Mpl z}  \right) e^{-\delta\Omega_{0}^{2}\left( (z-v_{g}^{+}t)^{2} + (z-v_{g}^{-}t)^{2}\right)}\\
&=& 2|A|^{2}e^{-2\delta\Omega_{0}^{2}(z-t)^{2}}\cos\left( 2\xi\frac{\Omega_{0}}{\Mpl}\Omega_{0}z\right)e^{ - 2\xi^{2}\frac{\Omega_{0}^{2}}{M^{2}}(\delta\Omega_{0}t)^{2}}\;.
\end{eqnarray}
This shows that there is an effect even on a linearly-polarised beam. The effect is a modulation of the wave intensity that depends quadratically on the energy and linearly on the distance of propagation. In addition, for a gaussian wave packet, there is a shift of the packet centre, that is controlled by the square of $\xi^{(3)}/\Mpl$ and hence is strongly suppressed with respect to the cosinusoidal modulation.

\subsection{Vacuum Birefringence}
\label{sec:birefringence}

The fact that electromagnetic waves with opposite ``helicities'' have slightly different group velocities, in EFT LV with $n=3$, implies that the polarisation vector of a linearly polarised plane wave with energy $k$ rotates, during the wave propagation over a distance $d$, through the angle \cite{Jacobson:2005bg}
\footnote{Note that for an object located at cosmological distance (let $z$ be its redshift), the distance $d$ becomes
\begin{equation}
d(z) = \frac{1}{H_{0}}\int^{z}_0 \frac{1+z'}{\sqrt{\Omega_{\Lambda} + \Omega_{m}(1+z')^{3}}}\,dz'\;,
\end{equation}
where $d(z)$ is not exactly the distance of the object as it includes a $(1+z)^{2}$ factor in the integrand to take into account the redshift acting on the photon energies.}
\begin{equation} 
\theta(d) = \frac{\omega_{+}(k)-\omega_{-}(k)}{2}d \simeq \xi^{(3)}\frac{k^2 d}{2\,M_{\rm Pl}}\;.
\label{eq:theta}
\end{equation} 

Observations of polarised light from a distant source can then lead to a constraint on $|\xi^{(3)}|$ that, depending on the amount of available information --- both on the observational and on the theoretical (i.e.~astrophysical source modeling) side --- can be cast in two different ways \cite{Maccione:2008tq}:
\begin{enumerate}
\item
Because detectors have a finite energy bandwidth, eq.~(\ref{eq:theta}) is never probed in real situations. Rather, if some net amount of polarization is measured in the band $k_{1} < E < k_{2}$, an order-of-magnitude constraint arises from the fact that if the angle of polarization rotation (\ref{eq:theta}) differed by more than $\pi/2$ over this band,
the detected polarization would fluctuate sufficiently for the net signal polarization to be suppressed \cite{Gleiser:2001rm, Jacobson:2003bn}.  
From (\ref{eq:theta}), this constraint is
\begin{equation} 
\xi^{(3)}\lesssim\frac{\pi\,M_{\rm Pl}}{(k_2^2-k_1^2)d(z)}\;,
\label{eq:decrease_pol}
\end{equation} 
%
This constraint requires that any intrinsic polarization (at source) not be
completely washed out during signal propagation. It thus relies on the
mere detection of a polarized signal; there is no need to consider the observed
polarization degree.
A more refined limit can be obtained by calculating the maximum
observable polarization degree, given the maximum intrinsic value \cite{McMaster}:
\begin{equation} 
\Pi(\xi) = \Pi(0) \sqrt{\langle\cos(2\theta)\rangle_{\mathcal{P}}^{2}
+\langle\sin(2\theta)\rangle_{\mathcal{P}}^{2}},
\label{eq:pol}
\end{equation} 
where $\Pi(0)$ is the maximum intrinsic degree of polarization,
$\theta$ is defined in eq.~(\ref{eq:theta}) and the average is
weighted over the source spectrum and instrumental efficiency,
represented by the normalized weight function
$\mathcal{P}(k)$~\cite{Gleiser:2001rm}.  
Conservatively, one can set $\Pi(0)=100\%$, but a lower value 
may be justified on the basis of source modeling.
Using \eqref{eq:pol}, one can then 
cast a constraint by 
requiring $\Pi(\xi)$ to exceed the observed value. 

\item
Suppose 
that
polarized light 
measured in a certain energy band 
has
a position angle $\theta_{\rm obs}$ with respect to a fixed
direction. At fixed energy, the polarization vector rotates by the
angle (\ref{eq:theta}) \footnote{Faraday rotation is negligible at
high energies.}; if the position angle is measured by averaging over a
certain energy range, the final net rotation 
$\left<\Delta\theta\right>$
is given by the
superposition of the polarization vectors of all the photons in that
range:
%
%
\begin{equation}
\tan (2\left\langle\Delta\theta\right\rangle) = \frac{
\left\langle\sin(2\theta)\right\rangle_{\mathcal{P}}}{\left\langle
\cos(2\theta)\right\rangle_{\mathcal{P}}}\;,
\label{eq:caseB}
\end{equation}
where 
$\theta$ is given by (\ref{eq:theta}).
If the position angle at emission $\theta_{\rm i}$ in the same energy band 
is known from a model of the emitting source, a constraint can be set by imposing
\begin{equation}
\tan(2\left\langle\Delta\theta\right\rangle) < \tan(2\theta_{\rm obs}-2\theta_{\rm i})\;.
\label{eq:constraint-caseB}
\end{equation}
%
Although this limit is tighter than those based on eqs.~(\ref{eq:decrease_pol}) and (\ref{eq:pol}), it clearly hinges on assumptions about the 
nature of the source, 
which may introduce significant uncertainties.
\end{enumerate}

The fact that polarised photon beams are indeed observed from distant objects imposes constraints on $\xi^{(3)}$. Recently, a claim of $|\xi^{(3)}| \lesssim 2 \times 10^{-7}$ was made using UV/optical polarisation measures from GRBs \cite{Fan:2007zb}. However, the strongest constraint to date comes from a local object. In \cite{Maccione:2008tq} the constraint $|\xi^{(3)}| \lesssim 6 \times 10^{-10}$ at 95\% Confidence Level (CL) was obtained by considering the observed polarization of hard-X rays from the Crab Nebula (CN) \cite{integralpol} (see also \cite{Forot:2008ud}).

\subsection{Threshold reactions}
\label{sec:thresholds}

An interesting phenomenology of threshold reactions is introduced by LV in EFT; also, threshold theorems can be rederived \cite{Mattingly:2002ba}. The conclusions of the investigation into threshold reactions are that \cite{Jacobson:2002hd}
\begin{itemize}
\item Threshold configurations still corresponds to head-on incoming particles and parallel outgoing ones
\item The threshold energy of existing threshold reactions can shift, and upper thresholds (i.e.~maximal incoming momenta at which the reaction can happen in any configuration) can appear
\item Pair production can occur with unequal outgoing momenta
\item New, normally forbidden reactions can be viable
\end{itemize}

LV corrections are surprisingly important in threshold reactions because the LV term (which as a first approximation can be considered as an additional mass term) should be compared not to the momentum of the involved particles, but rather to the (invariant) mass of the particles produced in the final state. Thus, an estimate for the threshold energy is
\begin{equation}
p_{\rm th} \simeq \left(\frac{m^{2}\Mpl^{n-2}}{\eta^{(n)}}\right)^{1/n}\;,
\label{eq:threshold-general}
\end{equation}
where $m$ is the typical mass of particles involved in the reaction.

Interesting values for $p_{\rm th}$ are discussed, e.g., in \cite{Jacobson:2002hd} and given in Tab.~\ref{tab:thresholds}. 
Reactions involving neutrinos are the best candidate for observation of LV effects, whereas electrons and positrons can provide results for $n=3$ theories but can hardly be accelerated by astrophysical objects up to the required energy for $n=4$. In this case reactions of protons can be very effective, because cosmic-rays can have energies well above 3 EeV.
\begin{table}[tbp]
\caption{Values of $p_{\rm th}$, according to eq.~(\ref{eq:threshold-general}), for different particles involved in the reaction: neutrinos, electrons and proton. Here we assume $\eta^{(n)} \simeq 1$.}
\begin{center}
\begin{tabular}{|c|c|c|c|}
\hline
& $m_{\nu}\simeq 0.1~\eV$ & $m_{e}\simeq 0.5~\MeV$ & $m_{p} \simeq 1~\GeV$ \\
\hline
$n=2$ & 0.1 eV & 0.5 MeV & 1 GeV\\
\hline 
$n=3$ & 500 MeV & 14 TeV & 2 PeV\\
\hline 
$n=4$ & 33 TeV & 74 PeV & 3 EeV\\
\hline
\end{tabular}
\end{center}
\label{tab:thresholds}
\end{table}%

\paragraph{$\gamma$-decay} The decay of a photon into an electron/positron pair is made possible by LV because energy-momentum conservation may now allow reactions described by the basic QED vertex. 
This process has a threshold that, if $\xi \simeq 0$ and $n=3$, is set by the condition
$k_{th} = (6\sqrt{3}m_e^2M/|\eta_\pm^{(3)}|)^{1/3}$ \cite{Jacobson:2005bg}. Furthermore, the decay rate is extremely fast above threshold \cite{Jacobson:2005bg}. The same conclusion holds when $n=4$. 

Because from birefringence $\xi \lesssim 9\times 10^{-10}$, the above expression for the photon decay can be used to constrain the electron/positron parameters. In \cite{Jacobson:2005bg} $|\eta_\pm| \lesssim 0.2$ was derived using the fact that 50~TeV $\gamma$-rays  from the CN were measured. This constraint has been tightened to $|\eta_\pm| \lesssim 0.05$, thanks to HEGRA's observations of 80~TeV photons \cite{Aharonian:2004gb}.

\paragraph{Vacuum \v{C}erenkov and Helicity Decay} In the presence of LV, the process of Vacuum \v{C}erenkov (VC) radiation $e^{\pm}\rightarrow e^{\pm}\gamma$ can occur. If we set $\xi \simeq 0$ and $n=3$, the threshold energy is given by
\begin{equation}
 p_{\rm VC} = (m_e^2M/2\eta^{(3)})^{1/3} \simeq 11~\TeV~\eta^{-1/3}\;.
\label{eq:VC_th}
\end{equation}
Just above threshold this process is extremely efficient, with a time scale of order $\tau_{\rm VC} \sim 10^{-9}~\s$ \cite{Jacobson:2005bg}. 

A slightly different version of this process is the Helicity Decay (HD, $e^{\mp}\rightarrow e^{\pm}\gamma$). If $\eta_{+} \neq \eta_{-}$, an electron can flip its helicity by emitting a suitably polarized photon. This reaction does not have a real threshold, but rather an effective one \cite{Jacobson:2005bg} --- $ p_{\rm HD} = (m_e^2M/\Delta\eta)^{1/3}$, where $\Delta\eta = |\eta_+^{(3)}-\eta_-^{(3)}|$ --- at which the decay lifetime $\tau_{HD}$ is minimized. For $\Delta\eta\approx O(1)$ this effective threshold is around 10 TeV.  
Note that below threshold $\tau_{\rm HD} > \Delta\eta^{-3} (p/10~\TeV)^{-8}\, 10^{-9}\s$, while above threshold $\tau_{\rm HD}$ becomes independent of $\Delta\eta$~\cite{Jacobson:2005bg}. 

\paragraph{Synchrotron radiation} Synchrotron emission is strongly affected by LV. In both LI and LV cases \cite{Jacobson:2005bg}, most of the radiation from an electron of energy $E$ is emitted at a critical frequency
\begin{equation}
 \omega_c = \frac{3}{2}eB\frac{\gamma^3(E)}{E} 
\label{eq:omega_sync}
\end{equation}
where $\gamma(E) = (1-v^2(E))^{-1/2}$, and $v(E)$ is the electron
group velocity. 
However, in the LV case, and assuming again $n=3$, the electron group velocity is given by
\begin{equation}
 v(E)= \frac{\partial E}{\partial p} =\frac{p}{E}\left(1+\frac{3}{2}\eta^{(3)}\frac{p}{M}\right)\,.
\end{equation}
Therefore, $v(E)$ can exceed $1$ if $\eta > 0$ or it can be strictly less
than $1$ if $\eta < 0$. 
This introduces a fundamental difference between particles with positive or negative
LV coefficient $\eta$. If $\eta$ is negative the group velocity of the electrons is strictly less than the (low energy) speed of light. This implies that, at sufficiently high energy, $\gamma(E)_{-} < E/m_e$, for all $E$. 
As a consequence, the critical frequency $\omega_c^{-}(\gamma, E)$ is always less than a maximal frequency $\omega_c^{\rm max}$ \cite{Jacobson:2005bg}.
Then, if synchrotron emission up to some frequency $\omega_{\rm obs}$ is observed, one can deduce that the LV coefficient for the corresponding leptons cannot be more negative than the value for which $\omega_c^{\rm max}=\omega_{\rm obs}$. Then, if synchrotron emission up to some maximal frequency $\omega_{\rm obs}$ is observed, one can deduce that the LV coefficient for the corresponding leptons cannot be more negative than the value for which $\omega_c^{\rm max}=\omega_{\rm obs}$, leading to the bound~\cite{Jacobson:2005bg}
\begin{equation}
\eta^{(3)}>-\frac{M}{m_e}\left(\frac{0.34\, eB}{m_e\,\omega_{\rm obs}}\right)^{3/2}\;.
\end{equation}

However, particles with positive LV coefficient can be superluminal. Therefore, at energies $E_c \gtrsim 8~\TeV /\eta^{1/3}$, $\gamma(E)$ begins to increase fasters than $E/m_e$ and reaches infinity at a finite energy, which corresponds to the threshold for soft VC emission. The critical frequency is thus larger than the LI one and the spectrum shows a characteristic bump due to the enhanced $\omega_c$.

\subsection{The Crab Nebula}
\label{sec:Crab}

All the knowledge described in the previous section about processes modified or simply allowed by LV can be used to infer properties of the radiation output of astrophysical objects and, eventually, to probe LV.
Amazingly, the Crab Nebula has proven an effective laboratory for such studies.

The CN is a source of diffuse radio, optical and X-ray 
radiation associated with a Supernova explosion observed 
in 1054~A.D. Its distance from Earth is approximately $1.9~\kpc$. 
A pulsar, presumably a remnant of the explosion, is located at the centre of
the Nebula. The Nebula emits an extremely broad-band spectrum (21 decades in frequency, see \cite{Maccione:2007yc} for a comprehensive list of relevant observations) that is produced by two major radiation mechanisms. 
The emission from radio to low energy $\gamma$-rays ($E < 1~\GeV$)
is thought to be synchrotron radiation from relativistic electrons, 
whereas inverse Compton (IC) scattering by these
electrons is the favored explanation for the higher energy 
$\gamma$-rays.  

From a theoretical point of view, the current understanding of the whole environment is based on the model presented in \cite{Kennel:1984vf}, which accounts for the general features observed in the CN spectrum.
We point the reader to \cite{Maccione:2007yc} for a wider discussion of this model.

Because we consider a LV version of electrodynamics, it is interesting to study whether this framework introduces modifications into the model of the CN, and if so, what effects it produces. How the synchrotron emission processes at work in the CN would appear in a ``LV world'' has been studied in \cite{Maccione:2007yc}. There the role of LV in modifying the characteristics of the Fermi mechanism (which is thought to be responsible for the formation of the spectrum of energetic electrons in the CN \cite{Kirk:2007tn}) and the contributions of VC and HD were investigated.

\paragraph{Fermi mechanism} Several mechanisms have been suggested for the formation of the spectrum of energetic electrons in the CN. As discussed in \cite{Maccione:2007yc}, the power-law spectrum of high energy ($> 1~\TeV$) particles is usually interpreted as arising from a first order Fermi mechanism operating at the ultra-relativistic termination shock front of the pulsar wind, because, in the simplest kinematic picture, this mechanism predicts a power law index of just the right value \cite{Kirk:2007tn}.
In \cite{Maccione:2007yc} the possible modifications affecting the Fermi mechanism due to LV have been discussed, including an interpretation of the high energy cut-off.
If we phenomenologically model the cut-off as an
exponential, then from the Fermi mechanism we would expect a particle spectrum in the high energy region $E>1\,$TeV, of the form $\displaystyle n(E)\propto\gamma(E)^{-p}\textrm{e}^{-E/E_{\rm c}}$ with $p\approx 2.4$ and $E_{\rm c}\approx 2.5\times10^{15}\,$eV. Then, we could safely deal with the electron/positron distributions inferred by \cite{Aha&Atoy}, making sure to replace the energy with the Lorentz boost factor.

\paragraph{Role of VC emission}  
The VC emission, due to its extreme rapidity above threshold, can produce a sharp cut-off in the acceleration spectrum. It has been verified that the modifications in the optical/UV spectrum produced by the VC radiation emitted by particles above threshold are negligible with respect to the synchrotron emission.

\paragraph{Role of Helicity Decay} 
To understand whether HD is effective, we must compare its typical time scale \cite{Jacobson:2005bg} with that of the spin precession of a particle moving in a magnetic field. From the discussion presented above, it is easy to see that $p_{\rm VC}$ is always slightly smaller than $p_{\rm HD}$, so that, for our purposes, the relevant regime of HD is that with momenta $p < p_{\rm HD}$. Thus, the typical time scale is $\tau_{\rm HD}\sim  10^{-9}~\s\times \Delta\eta^{-3} (p/10~\TeV)^{-8}$. Spin rotation effectively prevents the helicity decay if the precession rate is faster than the time needed for HD. Therefore, we can estimate that the HD becomes effective when the particle energy is above $p^{\rm (eff)}_{\rm HD} \gtrsim 930~\GeV \left(B/0.3~\mG\right)^{1/8}|\Delta\eta|^{-3/8}$.
Electrons and positrons with $E > p_{\rm HD}^{\rm (eff)}$ can be found only in the helicity state corresponding to the lowest value of $\eta_\pm$. Therefore, the population of greater $\eta$ is sharply cut off above threshold whereas the population with smaller $\eta$ is increased.

Using numerical tools previously developed \cite{Maccione:2007yc}, we can study the effect of LV ($n=3$) on the CN spectrum. This procedure requires that we fix most of the model parameters using radio to soft X-rays observations, which are not affected by LV \cite{Maccione:2007yc}. The high energy cut-off of the wind lepton spectrum $E_c \simeq 2.5~\PeV$ and a spectral index of the freshly accelerated electrons $p=2.4$ give the best fit to the data in the LI case \cite{Aha&Atoy}.

\begin{figure}[tbp]
 \centering
 \includegraphics[scale = 0.44, angle = 90]{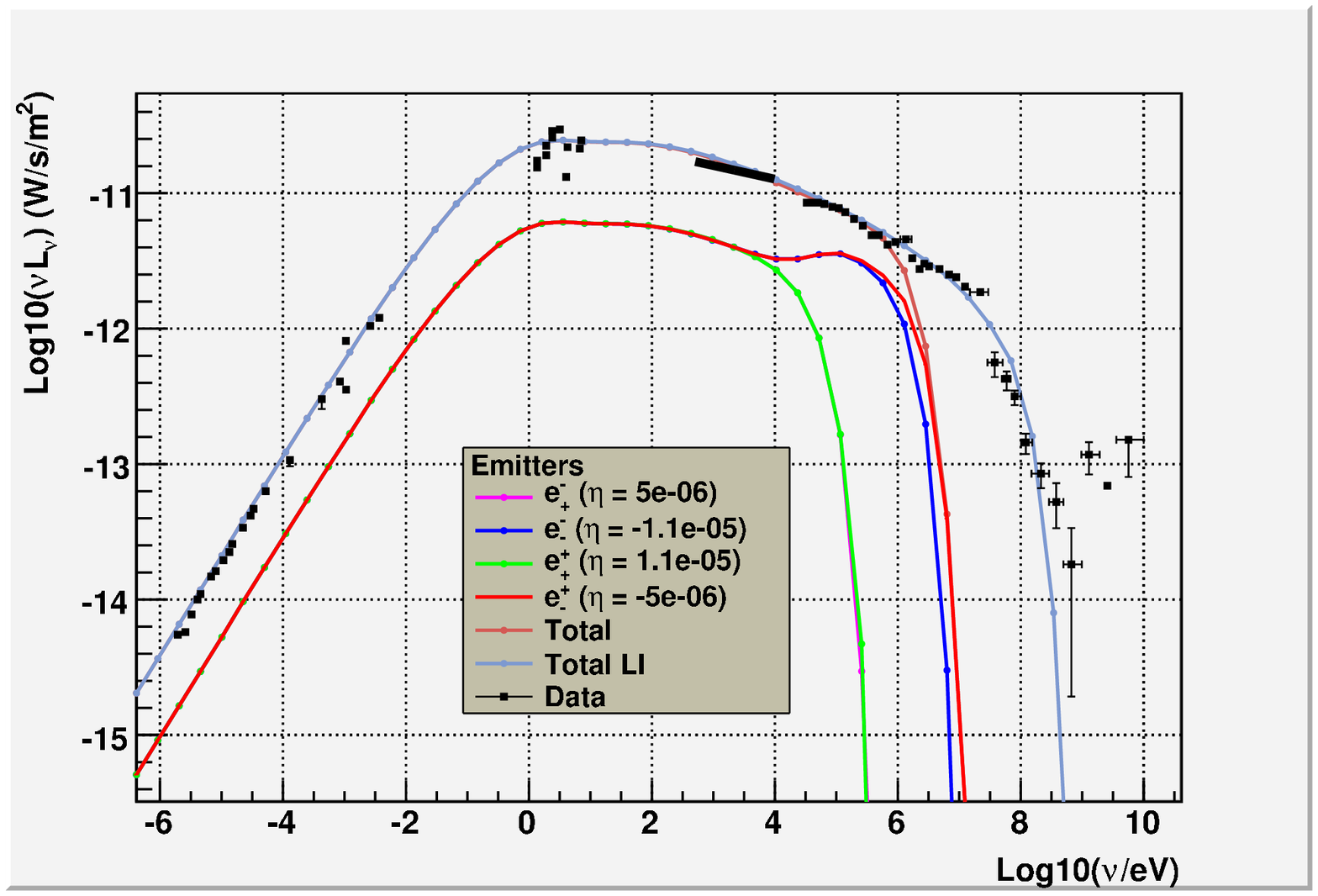}
 \includegraphics[scale = 0.44, angle = 90]{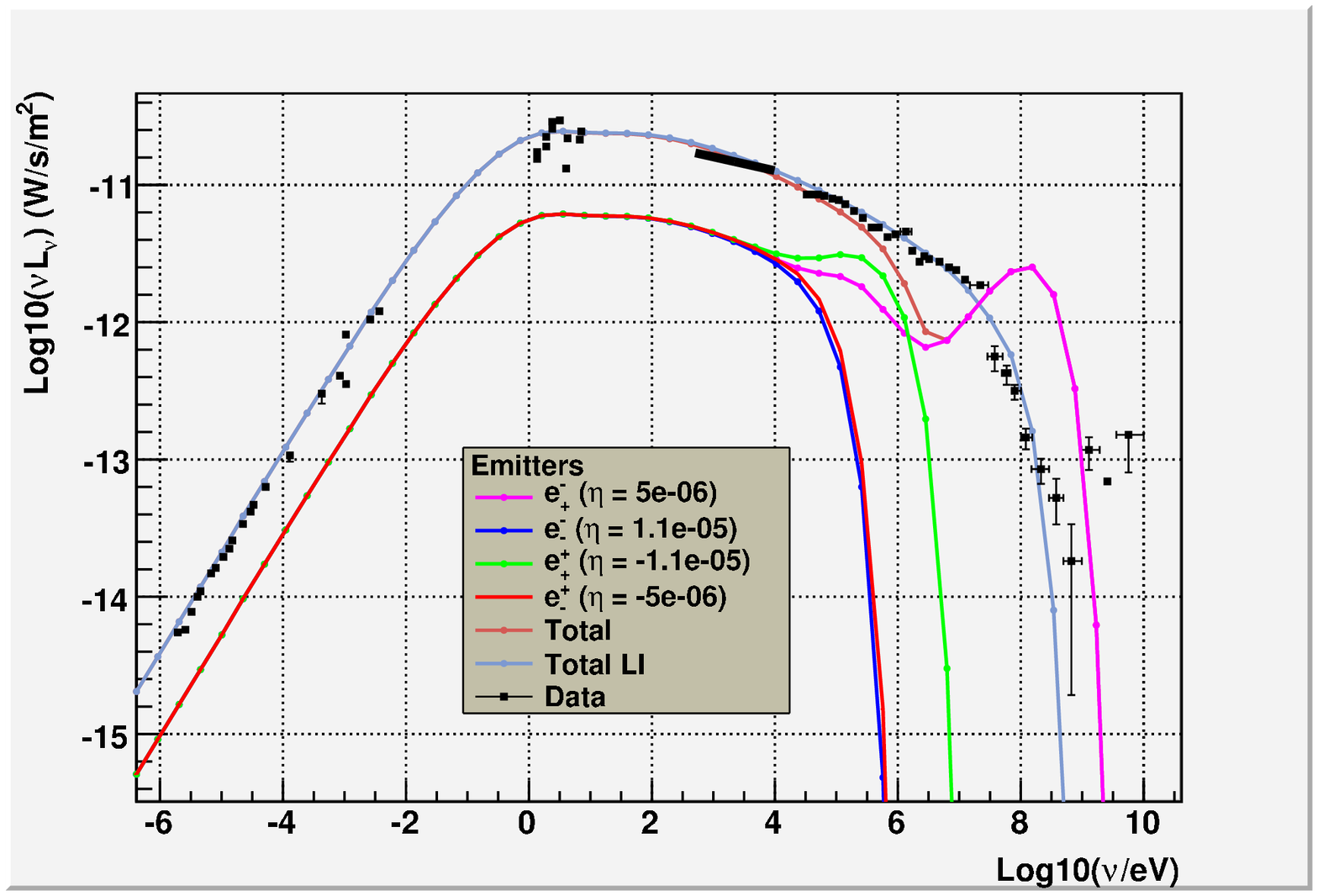}
 \caption{Comparison between observational data, the LI model and a LV
 one with $\eta_+\cdot\eta_- <0$ (left) and $\eta_+\cdot\eta_- >0$
 (right). The values of the LV coefficients, reported in the
 insets, show the salient features
 of the LV modified spectra. The leptons are injected according to the
 best fit values $p=2.4$, $E_c=2.5$ PeV. The individual contribution
 of each lepton population is shown.}
 \label{fig:spectra}
\end{figure}

Clearly only two configurations in the LV parameter space are truely different: $\eta_+\cdot\eta_- >0$ and $\eta_+\cdot\eta_- <0$, where $\eta_+$ is assumed to be positive for definiteness. The configuration wherein both $\eta_\pm$ are negative is the same as the $(\eta_+\cdot\eta_- >0,\,\eta_+>0)$ case, whereas that whose signs are scrambled is equivalent to the case $(\eta_+\cdot\eta_- <0,\,\eta_+>0)$. This is because positron coefficients are related to electron coefficients through $\eta^{af}_\pm = -\eta^{f}_\mp$ \cite{Jacobson:2005bg}. Examples of spectra obtained for the two different cases are shown in Fig.~\ref{fig:spectra}.

A $\chi^2$ analysis has been performed to quantify the agreement between models and data \cite{Maccione:2007yc}.
From this analysis, one can conclude that the LV parameters for the leptons are both constrained, at 95\% CL, to be $|\eta_\pm| < 10^{-5}$, as shown by the red vertical lines on the left-hand panel in Fig.~\ref{fig:constraints}. 
Although the best fit model is not the LI one, a careful statistical analysis (performed with present-day data) shows that it is statistically indistinguishable from the LI model at 95\% CL \cite{Maccione:2007yc}. 

\section{Constraints on $\bm{O(E/\Mpl)^{2}}$ Lorentz invariance violation}

The previous section attested to the strength of the constraints currently placed on dimension 5 LV operators. This is a remarkable achievement that was almost unforeseeable 10 years ago. However, it is true that the naturalness problem (see section \ref{sec:naturalness}) poses a challenge for the internal consistency of this approach to LV. Let us then move to the next order (mass dimension 6) LV operators and describe the status of the field. 

Ultra-High-Energy Cosmic Rays (UHECRs) have the potentiality to probe modified dispersion relations induced by CPT even dimension 6 operators. One of the most interesting features related to the physics of UHECRs is the Greisen-Zatsepin-Kuzmin (GZK) cut off \cite{Greisen:1966jv,1969cora...11...45Z}, a suppression of the high-energy tail of the UHECR spectrum arising from interactions with CMB photons, according to $p\gamma\rightarrow \Delta^{+}\rightarrow p\pi^{0}(n\pi^{+})$. This process has a (LI) threshold energy $E_{\rm th} \simeq 5\times 10^{19}~(\omega_{b}/1.3~\meV)^{-1}~\eV$ ($\omega_{b}$ is the target photon energy). Experimentally, the presence of a suppression of the UHECR flux was confirmed only recently \cite{Abbasi:2007sv,Roth:2007in}. Although the cut off could be also due to the finite acceleration power of the UHECR sources, the fact that it occurs at the expected energy favors the GZK explanation. The results shown in \cite{Cronin:2007zz} further strengthen this hypothesis.

As a threshold process, photopion production is strongly affected by LV. Several authors have studied the constraints implied by the detection of such a suppression \cite{Aloisio:2000cm,Alfaro:2002ya,Jacobson:2002hd,Mattingly:2008pw,Scully:2008jp,Stecker:2009hj}. However, a detailed LV study of the GZK feature is hard to perform, because of the many astrophysical uncertainties related to the modeling of the propagation and the interactions of UHECRs.

Rather surprisingly, however, significant limits on $\xi$ and $\eta$ can be derived by considering UHE photons \cite{Galaverni:2007tq,Maccione:2008iw}, further improving the constraints on dimension 5 LV operators and providing a first robust constraint of QED with dimension 6 CPT even LV operators. UHE photons originate in the interactions of UHECRs with the CMB (GZK process), leading to the production of neutral pions that subsequently decay into photon pairs. These photons are mainly absorbed by pair production onto the CMB and radio background. Thus, the fraction of UHE photons in UHECRs is theoretically predicted to be less than 1\% at $10^{19}~\eV$ \cite{Gelmini:2005wu}. Several experiments imposed limits on the presence of photons in the UHECR spectrum. In particular, the photon fraction is less than 2.0\%, 5.1\%, 31\% and 36\% (95\% C.L)~at $E = 10$, 20, 40, 100 EeV  respectively \cite{Aglietta:2007yx,Rubtsov:2006tt}. 

However, pair production is strongly affected by LV. In particular, the (lower) threshold energy can be slightly shifted and in general an upper threshold can be introduced \cite{Jacobson:2002hd}. If the upper threshold energy is lower than $10^{20}~\eV$, then UHE photons are no longer attenuated by the CMB and can reach the Earth, constituting a significant fraction of the total UHECR flux and thereby violating experimental limits \cite{Galaverni:2007tq,Maccione:2008iw,Galaverni:2008yj}. 

Moreover, it has been shown \cite{Maccione:2008iw} that the $\gamma$-decay process can also imply a significant constraint. Indeed, if some UHE photon ($E_{\gamma}\simeq 10^{19}~\eV$) is detected by experiments (and the Pierre Auger Observatory, PAO, will be able to do so in few years \cite{Aglietta:2007yx}), then $\gamma$-decay must be forbidden above $10^{19}~\eV$. 

\begin{figure}[tbp]
 \includegraphics[scale=0.44, angle = 90]{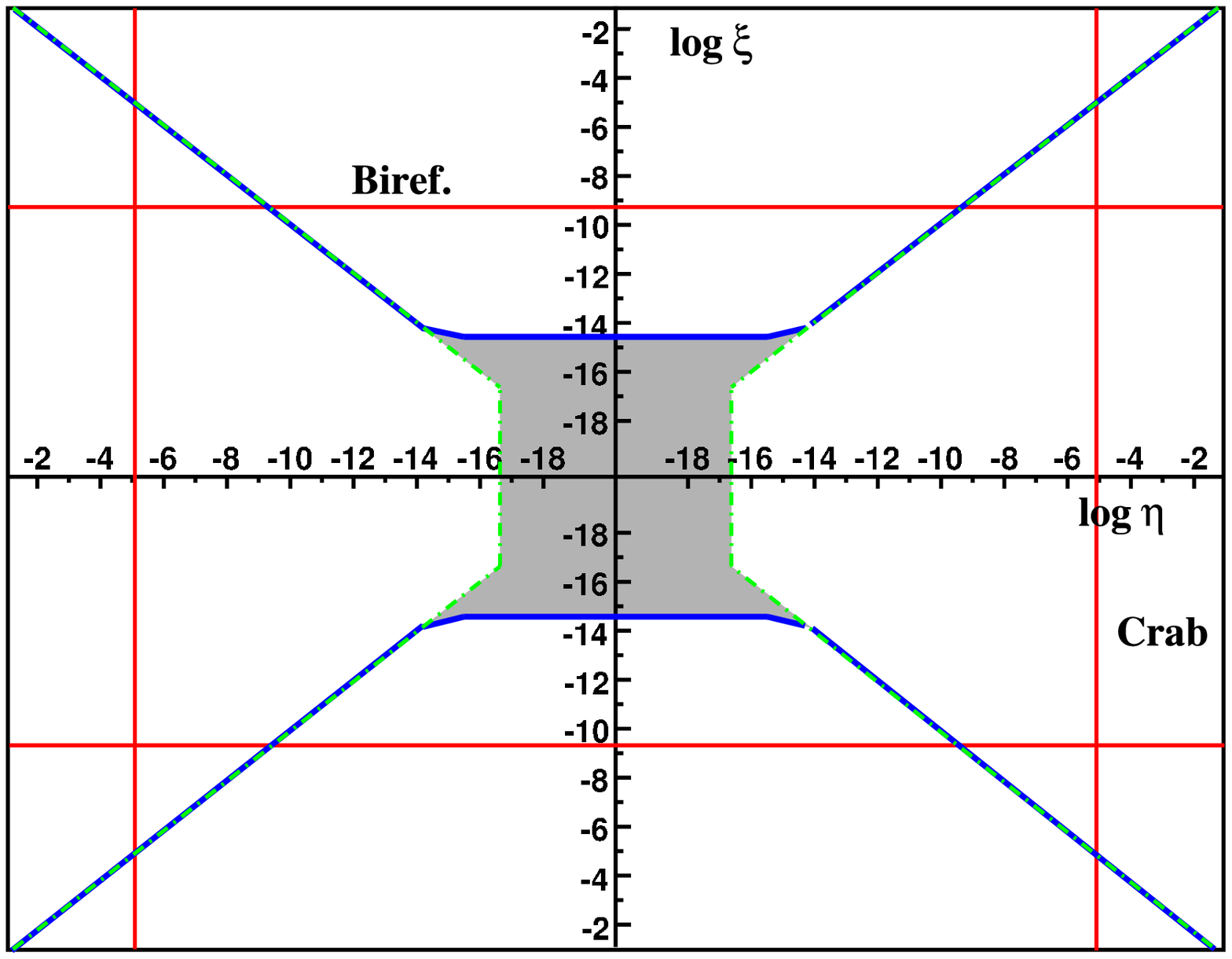}
 \includegraphics[scale=0.45, angle = 90]{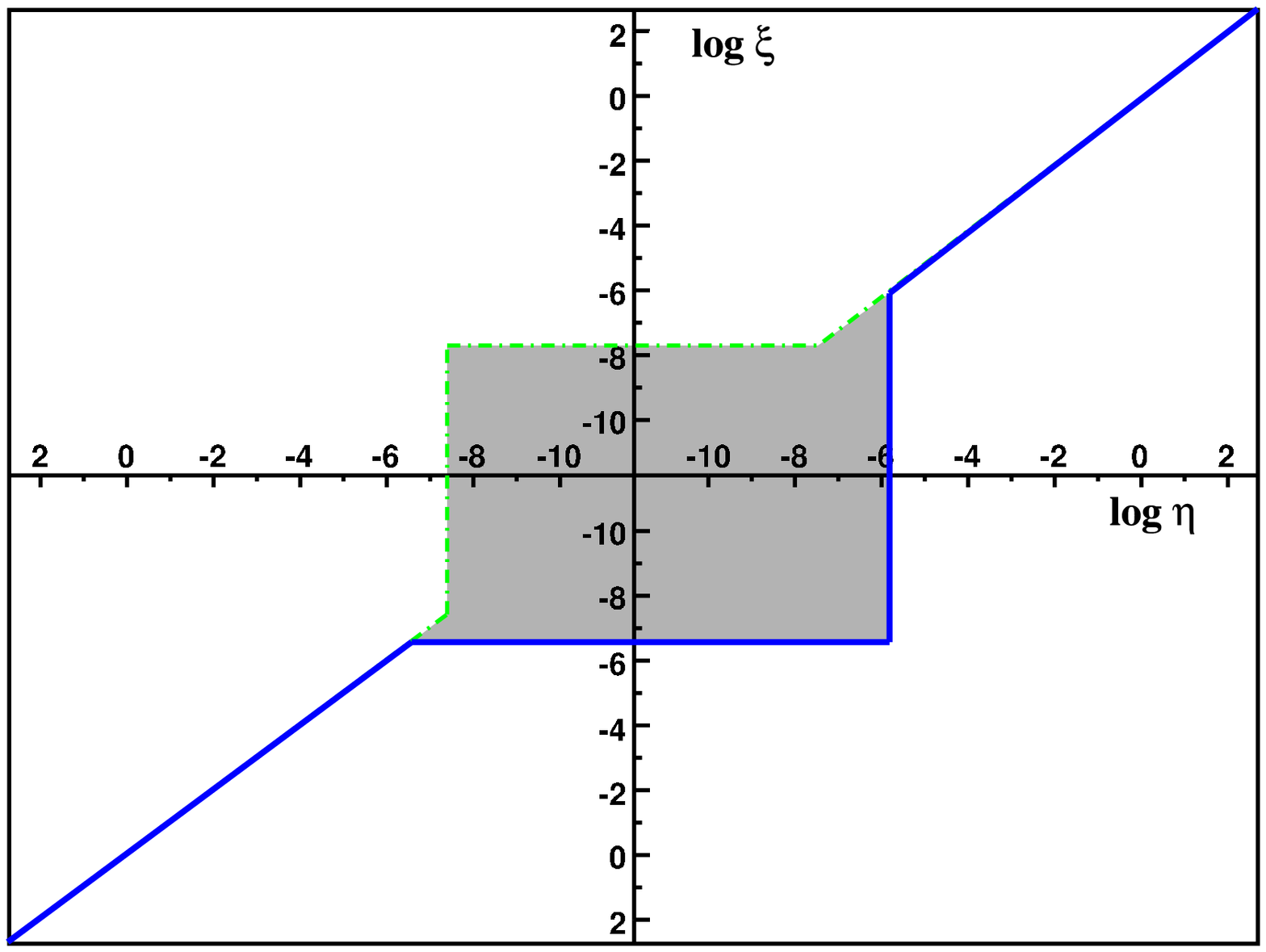}
 \caption{Left panel: LV induced by dimension 5 operators. Right panel: LV induced by dimension 6 operators. The LV parameter space is shown. The allowed regions are shaded grey. Green dotted lines represent values of $(\eta, \xi)$ for which the $\gamma$-decay threshold $k_{\gamma-dec} \simeq 10^{19}~\eV$. Solid, blue lines indicate pairs $(\eta,\xi)$ for which the pair production upper threshold $k_{\rm up} \simeq 10^{20}~\eV$. Red lines correspond to constraints obtained by Crab Nebula observations.}
 \label{fig:constraints}
\end{figure}

In conclusion we show on the left-hand panel of Fig.~\ref{fig:constraints} the overall picture of the constraints of dimension 5 and 6 LV operators, where the green dotted lines do not correspond to real constraints, but to the ones that will be achieved when AUGER will observe, as expected, some UHE photon.

\section{Conclusions}

We hope that this review has convinced even the most skeptical reader that it is now possible to strongly constrain Planck-suppressed effects motivated by QG scenarios. The above discussion makes clear that this can be achieved because even tiny violations of a fundamental symmetry such as Lorentz invariance can lead to detectable effects at energies well below the Planck scale. Although there are several proposed frameworks to deal with QG induced LV, we have discussed here the most conservative framework of LV in EFT, given its simplicity and robustness with respect to the eventual UV completion of the theory (although there are also QG scenarios that do not admit an EFT low-energy limit). Given that there is a rich literature about experimental constraints on EFT with renormalizable LV operators \cite{Kostelecky:2008zz}, we focused here on astrophysical constraints on high energy Planck-suppressed LV induced by non-renormalizable operators of mass dimension 5 and 6.\footnote{It was recently recognized that the SM with some LV higher mass dimension operators can be renormalizable \cite{Anselmi:2007ri,Anselmi:2008bq,Anselmi:2008bs,Anselmi:2008bt,Visser:2009fg,Anselmi:2009vz}. It might be then interesting to further investigate the phenomenological relevance of these models.} 

The outcome of this investigation is summarized in Fig.~\ref{fig:constraints} which clearly shows that dimension 5 LV operators in QED are extremely constrained to the level of convincingly ruling them out. This may be an observational support to the theoretical issue regarding the naturalness of such operators (discussed in section \ref{sec:naturalness}). Thus, it is interesting that the theoretically favoured dimension 6, CPT even, LV operators in QED are constrained by UHECR observations. Although to date we only have limits implied by the non-observation of a significant photon fraction in the UHECR flux above $10^{19}~\eV$, the expected detection of some UHE photon by the PAO will further restrict the allowed region of parameter space for LV QED to the portion of the grey region in Fig.~\ref{fig:constraints} limited by the green dotted line and including the origin.

The strength of the constraints achieved so far is a strong indication either that Lorentz invariance may be an exact symmetry of Nature or that the framework incorporating departures from it cannot be cast in EFT form. Therefore, it is worthwhile to consider alternative scenarios, such as those presented in section~\ref{sec:others}. Given that some of these alternative frameworks currently do not allow full control of the dynamics, one must look back at purely kinematical (e.g.~time-of-flight) observations. For example, the $D-$brane inspired model discussed in section~\ref{sec:nonEFT} leads to a non-birefringent dispersion relation for photons, with $O(E/\Mpl)$ LV, thereby evading birefringence constraints. However, such dispersion relations can be probed by time-of-flight, as discussed in section \ref{subsec:tof}.


Interestingly, the authors of \cite{Albert:2007qk} discussed LV as a possible explanation of the arrival time structure of TeV photons from Mkn 501, determining that the best-fit to data may be obtained by adopting $\xi \sim O(1)$. Quite intriguingly, a similar value seems to be compatible \cite{Ellis:2009yx} with the limits imposed by the recent time delay observations made by the HESS collaboration \cite{Aharonian:2008kz} of the AGN PKS 2155-304 and by the FERMI satellite of the gamma-ray burst GRB080916 \cite{2009Sci...323.1688A}. If this were the case --- and if the more probable astrophysical explanations for these delays could be reasonably excluded --- then we would have to admit that the EFT description of LV phenomena related to QG would fail at TeV energies. In fact, the given best-fit value of $\xi$ would exceed the best constraint on $\xi$ in EFT by several orders of magnitude \cite{Maccione:2008tq}. This fact would reveal a very peculiar feature of the underlying QG theory, hence strongly restricting suitable scenarios.

We are still at a preliminary stage in casting constraints on non-QED sectors of the SM. In particular, although order-of-magnitude constraints have been obtained on high energy LV in the hadronic sector, a complete reconstruction of the UHECR spectrum recently led to more robust constraints, which are as strong as $10^{-3}$ and $10^{-1}$ for dimension 6 LV coefficients for protons and pions, respectively \cite{Maccione:2009ju}. Furthermore, the neutrino sector is still largely unexplored, in spite of the very low energies at which threshold effects should be affected by high energy LV in EFT. This is mainly due to the difficulty of theoretically reconstructing and experimentally observing neutrino spectra at high energy. Still, one may hope that planned future experiments in neutrino astrophysics may eventually provide the much needed information to further explore this SM sector.

As we discussed at the beginning of this review, LV is not the only possible low energy QG signature. Nonetheless, it is encouraging that it was possible to gather such strong constraints on this phenomenology in only a few years. This should motivate researchers to further explore this possibility as well as to look even harder for new QG induced phenomena that will be amenable to observational tests.
This will not be an easy task, but the data so far obtained prove that the Planck scale is not so untestable after all.

\section*{Acknowledgements}

LM acknowledges support from SISSA, where a substantial part of the work included in this review was done.

\bibliographystyle{apsrev}	

\end{document}